\documentclass[acmsmall,screen, nonacm]{acmart}

\AtBeginDocument{%
  }

\setcopyright{acmlicensed}
\copyrightyear{2018}
\acmDOI{XXXXXXX.XXXXXXX}
\begin{document}

\title{Rigorous Evaluation of Microarchitectural Side-Channels with Statistical Model Checking}


\author{Weihang Li}
\affiliation{
  \institution{Duke University}
  \country{USA}
}
\email{weihang.li@duke.edu}
\author{Pete Crowley}
\affiliation{%
  \institution{Duke University}
  \country{USA}
}
\email{pete.crowley@duke.edu}
\author{Arya Tschand}
\affiliation{%
  \institution{Harvard University}
  \country{USA}
}
\email{aryatschand@g.harvard.edu}
\author{Yu Wang}
\affiliation{%
  \institution{University of Florida}
  \country{USA}
}
\email{yuwang1@ufl.edu}
\author{Miroslav Pajic}
\affiliation{%
  \institution{Duke University}
  \country{USA}
}
\email{miroslav.pajic@duke.edu}
\author{Daniel Sorin}
\affiliation{%
  \institution{Duke University}
  \country{USA}
}
\email{sorin@ee.duke.edu}

\begin{abstract}
Rigorous quantitative evaluation of microarchitectural side channels is challenging for two reasons. First, the processors, attacks, and defenses often exhibit probabilistic behaviors. These probabilistic behaviors arise due to natural noise in systems (e.g., from co-running processes), probabilistic side channel attacks, and probabilistic obfuscation defenses.
Second, microprocessors are extremely complex. Previous evaluation methods have relied on abstract or simplified models, which are necessarily less detailed than real systems or cycle-by-cycle simulators, and these models may miss important phenomena.  Whereas a simple model may suffice for estimating performance, security issues frequently manifest in the details. 

We address this challenge by introducing Statistical Model Checking (SMC) to the quantitative evaluation of microarchitectural side channels.  SMC is a rigorous statistical technique that can process the results of probabilistic experiments and provide statistical guarantees, and it has been used in computing applications that depend heavily on statistical guarantees (e.g., medical implants, vehicular computing). With SMC, we can treat processors as opaque boxes, and we do not have to abstract or simplify them. We demonstrate the effectiveness of SMC through three case studies, in which we experimentally show that SMC can evaluate existing security vulnerabilities and
defenses and provide qualitatively similar conclusions with greater statistical rigor, while making no simplifying assumptions or abstractions.  We also show that SMC can enable a defender to quantify the amount of noise necessary to have a desired level of confidence that she has reduced an attacker’s probability of success to less than a desired threshold, thus providing the defender with an actionable plan for obfuscation via noise injection.
\end{abstract}

\keywords{Microarchitecture, Security, Side-channel attack, Experimental Methodology}


\maketitle

\section{Introduction}

While some security attacks have deterministic outcomes---that is, the attack works every time (or fails every time)---many security issues are inherently probabilistic.\footnote{Formally speaking, some of the behaviors we consider are non-deterministic (i.e., there may not exist a predefined probability distribution governing them) rather than probabilistic. 
Still, to simplify our presentation, we will refer to all such behaviors as being probabilistic.}
Particularly for side-channel attacks, which are the focus of our work, probabilistic behavior is difficult to avoid because of four underlying reasons.  First, most systems naturally have ``noise'' in their performance; if you run a program multiple times on a given computer, it will take a different latency (and energy consumption) each time.  This natural noise is a result of other processes and differences in microarchitectural state (e.g., the state of the branch predictor). Second, some attacks are probabilistic; the attack may be run many times (e.g., by requesting encoding of different plaintext inputs) and the secret is decoded by statistically processing the results of the multiple runs.  Third, some defenses are probabilistic; a victim program may try to obfuscate side channels by injecting ``noise'', and this noise is often probabilistic.  Fourth, the actual act of measurement (of latency, energy, etc.) can itself be noisy.

In the presence of probabilistic behaviors, it can be difficult to quantitatively evaluate the security.  We use an example here, plus several more later, to highlight the types of security questions we would like to answer but cannot with current methodologies.

\vspace{0.2cm}

\noindent \underline{Motivating Example}: You are running software that is known to be vulnerable to a probabilistic side-channel attack. You can reduce the probability of the attack being successful via obfuscation, i.e., injecting "noise" into the system (e.g., via extra memory accesses to pseudo-random addresses).  You know that injecting more noise reduces the probability of attack success but it also causes more performance loss, so you do not want to inject more noise than necessary. You can also thwart the attacker by reconfiguring the software (e.g., changing a cryptographic seed), but the performance loss from reconfiguration limits you to doing it no more than once per minute.
What is the minimum amount of noise you should inject to have 99\% confidence that the attacker will have less than a 1\% chance of success within one minute of attack time?  You would like to know the answer to that question, but we currently do not have the methodological tools to discover it.

In this work, we develop a methodology for security evaluation that enables us to answer questions like that, and it is based on the following goals. 

\vspace{0.1cm}

\vspace{0.1cm}

\noindent\textbf{Avoid making assumptions about probabilistic behaviors}.  It can be tempting for a computer architect to model probabilistic behaviors with probabilistic models, but such models generally must make assumptions if they are to be tractable.  For example, one common assumption is to assume that a form of noise has a Gaussian distribution.  More generally, probabilistic models make assumptions about distributions and relationships (e.g., two random variables are independent).  While these assumptions may often be true or ``close enough,'' we 
ideally prefer not to have to make any assumptions.  Thus we could be confident in our results, without having to worry about the validity of our assumptions in every situation.  

\vspace{0.1cm}

\noindent\textbf{Avoid relying on over-simplified or abstract models}.  Computer architects develop models for studying systems that do not (yet) exist.  Often these models are highly detailed simulators that capture the intricate behaviors of the processors being studied.  Architects will use more abstract models (e.g., queuing models) for initial studies or design space exploration, but when we really need to be sure, we turn to more detailed models.  In security, there has been a significant amount of work in models, both probabilistic~\cite{deng:hasp:2018,deng:jhss:2019,deng:ieeetc:2021, yang:isca:2023} and formal~\cite{deutsch:isca:2023,domnitser:cns:2010,bourgeat:micro:2020, genkin:asiaccs:2023,he:micro:2017}, but the models are often far more simple or abstract than the processors under study to enable tractable analysis.  Because security issues often manifest in the details, we want a methodology that can accommodate sophisticated models like cycle-accurate simulators.

\vspace{0.1cm}

\noindent\textbf{Provide statistical rigor}.  When working with probabilistic behavior, our results are necessarily probabilistic.  For example, we may run an attack multiple times with different outcomes (e.g., measured side channel latencies).  To properly process these samples, we must use rigorous mathematical methods.  We seek to provide statistical confidence for our results. 

\vspace{0.1cm}


To meet all of these goals, we propose the use of statistical model checking (SMC) \cite{legay2010statistical, agha2018survey} for evaluation of architectural security.  SMC is a well-known methodology for evaluating systems with probabilistic behaviors, and
it has been applied to cyber-physical systems that often need to provide statistical guarantees~\cite{zarei_hscc20,wang_tecs19}. For example, a medical device might need to achieve 99.9\% confidence that its embedded processor will never be in a certain state for longer than a specified threshold of time.  More recently, SMC has been applied to the experimental evaluation of microarchitecture, particularly its performance in the presence of variability (e.g., due to the microarchitectural state in which a core begins executing a program)~\cite{mazurek:micro:2023}.  SMC treats the system under study as an ``opaque box'', requiring no model assumptions or simplifications; it is compatible with studies performed on cycle-by-cycle simulators and real systems.  It is based on rigorous mathematics, and it can evaluate sophisticated properties (e.g., properties expressed in linear temporal logic). 

In this work, we make the following contributions:

\begin{itemize}
    \item We demonstrate SMC's ability to provide statistically rigorous evaluation of security for systems with probabilistic behaviors, and we do so using results from a standard, highly-detailed simulator -- there is no need to build new probabilistic or formal models.
    \item We develop a set of expressive properties that provide insight into system security and that can be used in future security evaluations.
    \item Through three case studies, we experimentally show that SMC can evaluate existing security vulnerabilities and defenses and provide qualitatively similar conclusions with greater statistical rigor, while making no simplifying assumptions or abstractions.  
    \item We show that SMC can enable a defender to quantify the amount of noise necessary to have a desired level of confidence that she has reduced an attacker's probability of success to less than a desired threshold, thus providing the defender with an actionable plan for obfuscation via noise injection.

\end{itemize}

\hyphenpenalty=10000
\exhyphenpenalty=10000

\section{Assumptions}
\label{sec:assumptions}

We make the following assumptions in this work:

First, the system, attacker, or defense exhibit probabilistic behaviors. (These behaviors do not, however, need to adhere to any particular or predefined probability distribution.)  In the presence of probabilistic behaviors, we seek to understand averages and distributions of results rather a single, worst-case outcome.  

Second, the attacker's strategy is fixed, i.e., does not change algorithmically over the course of the attack.  Imagine that the attacker's actions can be modeled with a flowchart.  The attacker can follow different paths through the flowchart based on what she observes during the attack, but she cannot switch to a different flowchart.  The mathematical foundation of SMC assumes that the underlying system can be modeled as a Markov Chain (which requires a fixed strategy), but not a Markov Decision Process (which accommodates a changeable strategy).  It is critical to note that we do not need to construct a Markov Chain, though, to use SMC.

Third, we are evaluating known attacks.  We do not claim that SMC can be used to make claims about unknown attacks.
\section{Tutorial on Statistical Model Checking for Processor Analysis}

SMC is a rigorous statistical 
approach used to formally evaluate performance properties of opaque-box systems that exhibit unknown variability. Recently, it has been successfully adapted to assess computer system performance from arbitrary result distributions with probabilistic guarantees \cite{mazurek:micro:2023}. SMC can be applied to sample results from either hardware or simulator experiments with unknown probabilistic models, and it avoids the limitations of prior approaches, such as assuming Gaussian distributions for unknown variability \cite{alameldeen:hpca:2003}. 

\subsection{Overview}

From a statistical point of view, a system's evaluation metric $X$ (e.g., runtime) is a random variable whose value varies across repeated experiments due to system variability. Because of this variability, the probability distribution $P_X$ of $X$ can differ significantly across computer systems and software workloads. Ideally, to fully capture $X$, we would present the entire probability distribution $P_X$ with all possible values of $X$ in various scenarios. However, this process requires a large number of experiments, which is infeasible in most cases. As a result, in practice, we seek to answer questions about certain statistical characteristics of the probability distribution.

SMC statistically asserts \emph{properties} of a computer system, where properties are binary functions involving one or more evaluation metrics. A property is either true or false for a given system execution, and it may vary among executions due to system variability. SMC can answer questions about the probability that a computer satisfies these properties, i.e., whether a property $\varphi$ (which is related to the behavior of a computer system) will hold for at least an $F \in [0, 1]$ fraction of executions. SMC is preferable to estimating the expected value for two key reasons, discussed below. 
In general, compared to existing methods, SMC has several advantages that make it ideal for studying security problems in computer systems.

First, SMC can handle properties that describe arbitrarily complex situations, as long as they can be expressed in a formal symbolic language such as temporal logic. For instance, SMC can be used to evaluate randomization-based defenses against side-channel attacks by assessing whether, at a given noise level, the attacker's accuracy exceeds a specified threshold. Such properties can be formulated in temporal logic and evaluated using SMC. Additional examples are discussed in our three case studies.

Moreover, SMC does not require assumptions about the probability distribution of the evaluation metric. Previous approaches (e.g., \cite{alameldeen:hpca:2003})
often assume a Gaussian distribution for $P_X$ and focus on estimating the expected value $\mathbb{E}[X]$ by averaging $N$ sample executions, as the obtained $\bar{X}$ converges to $\mathbb{E}[X]$ with large $N$. However, estimating the error accurately for small $N$ when $P_X$ is unknown is challenging. In simulation studies, $N$ is typically small (around 3-5~\cite{carlson:sc:2011}) and some prior work (e.g., \cite{alameldeen:hpca:2003}) that assumes that $\bar{X}$ follows a Gaussian distribution may not always be valid. More discussion can be found in Mazurek et al.~\cite{mazurek:micro:2023}. 

Another key advantage of SMC is its ability to perform rigorous statistical inference even when the underlying randomness is unknown. Since the truth value of properties is binary, SMC can directly leverage statistical inference techniques for binomial distributions to provide robust guarantees. Specifically, given $N$ sample executions from independent experiments, SMC can determine not only whether the probability of satisfying a property $\varphi$ exceeds a threshold $F$, but also compute the confidence level $C \in (0, 1)$ of this result. The confidence level ensures that \emph{for at least a fraction $C$ of all possible sets of $N$ random executions, our answer matches the ground truth}.

\subsection{Technical Details}

Mathematically, the basic problem of SMC is to check if a property $\varphi$ holds on a computer system $S$ with probability greater than a proportion $F$:  
\begin{equation} \label{eq:goal}
p_\varphi := \mathbb{P}_{\sigma \sim S} (\varphi \text{ holds on } \sigma) \geq F,
\end{equation}
where $\sigma$ is a random execution of the computer system $S$ with probabilistic behavior, $\sigma \sim S$ indicates that $\sigma$ is ``drawn'' from the system~$S$, and $\varphi$ is the given property of interest. Effectively, we want to know if the probability $p_\varphi$ that an execution $\sigma$ of the system $S$ satisfies the property $\varphi$ is greater than $F$. Although SMC can handle more complex statements, we focus on \eqref{eq:goal} because it can cover most evaluation questions for computer systems.  While we consider multiple values of $F$, we note that $F=0.5$ is often of intuitive interest in that it represents the median.

For computer security, common properties of interest are expressible in signal temporal logic (STL) \cite{maler2004monitoring}. 
All STL formulas have well-defined semantics; they can be parsed and yield a meaning without ambiguity~\cite{maler2004monitoring}. This feature guarantees that SMC will never ``misunderstand'' a property specified in STL. For simplicity, we can view properties as binary random variables taking values $true$ or $false$ for each sample execution. Such truth values can be derived formally by computer algorithms by a standard process, for which we refer the readers to the literature ~\cite{maler2004monitoring} and subsequent work. 

SMC differs from 
(symbolic) model checking in estimating the probability $p_\varphi$ in~\eqref{eq:goal} based on the true values of $\varphi$. When it is infeasible to derive a probabilistic model of the system, as is the case for many complex security problems, computing the probability from model knowledge as in model checking is impossible. Instead, SMC employs statistical inference to estimate $p_\varphi$ from a finite set of samples as follows. 
Consider a set of $N$ sample executions $\sigma_1, \ldots, \sigma_N$ taken independently from repeated experiments on the system $S$. For $i = 1,\ldots,N$, with a slight abuse of notation,~let
\begin{equation} \label{eq:varphi}
\varphi(\sigma_i) = \begin{cases}
1, & \text{if $\varphi$ is true on } \sigma_i, \\
0, & \text{otherwise.}
\end{cases}
\end{equation}

The total number of executions satisfying $\varphi$, denoted by $M = \sum\nolimits_{i \in [N]} \varphi ( \sigma_{i})$, follows a binomial distribution $\mathrm{Binom}(N, p_\varphi)$, where $p_\varphi$ is the probability that $\varphi$ is true on a given execution. The statistic $M/N$ is an unbiased estimator for $p_\varphi$. Intuitively, when $M/N < F$, we assert \textit{negative} for the condition in~\eqref{eq:goal}; otherwise, we assert \textit{positive}. Thus, the statistical assertion based on the $N$ sample executions is defined as
\begin{equation} \label{eq:assert_simple}
\mathcal{A} ( \sigma_1,\ldots,\sigma_N ) = 
\begin{cases}
\text{negative}, & \text{if } M/N < F, \\
\text{positive}, & \text{if } M/N \geq F.
\end{cases}
\end{equation}

Due to the randomness of sample executions, the statistical assertion $\mathcal{A}$ from Equation~\eqref{eq:assert_simple} does not always align with the ground truth. The statistical accuracy of $\mathcal{A}$ is captured by its confidence level $C$, which ensures that
\begin{align*}
\mathbb{P}_{\sigma_1, \ldots \sigma_N \sim S} \big( \mathcal{A} \text{ is negative/positive} \ | \ & \text{Condition \eqref{eq:goal}} \text{ is true/false} \big)  \leq 1 - C
\end{align*}
This means that for at least a fraction $C$ of all possible sets of sample executions, the value of $\mathcal{A}$ matches the ground truth. For example, if $C = 0.99$, we expect at most one disagreement after using the statistical assertion $100$ times (i.e., on the results of $100$ sample executions). We also refer to $1 - C$ as the \textit{significance level}, which can be interpreted as the maximum of the Type I or Type II error rates (or false positive/negative rates) from statistical literature.

The confidence level of $\mathcal{A}$ is computed using the Clopper-Pearson exact method:
\begin{align}
\label{eq:alpha_cp}
& C_{\text{CP}} (a, b \, \vert \, M, N) = 
\\ & \quad 
\begin{cases}
(1-a)^{N} - (1-b)^{N},
& \text{ if } M = 0 \\
b^{N} - a^{N},
& \text{ if } M = N \\
\mathcal{B} ( b \, \vert \, M + 1, N - M ) - \mathcal{B} ( a \, \vert \, M, N - M + 1 ),
& \text{ else;}
\end{cases}\nonumber
\end{align} 
where $\mathcal{B}(\cdot \mid x_1, x_2)$ is the cumulative distribution function (CDF) of the beta distribution with shape parameters $(x_1, x_2)$. The values of $a$ and $b$ are defined as:
\begin{equation} \label{eq:alpha_cp2}
\begin{cases}
a = 0, \ b = F, & \text{if } M/N < F, \\
a = F, \ b = 1, & \text{if } M/N \geq F.
\end{cases}
\end{equation}
This method is statistically accurate for any sample size, unlike most previous statistical methods for computer systems, which are only asymptotically accurate~\cite{kalibera:arxiv:2007,irving:fcs:2020,chen:hpca:2012}. Moreover, the Clopper-Pearson method is specifically designed for the binomial random variable $M$ and achieves optimal sample efficiency among statistically accurate methods~\cite{clopper1934use}.

SMC can be designed to achieve any desired confidence level $C \in (0,1)$ by running in a loop. It draws new sample executions and updates the confidence level $C_{\text{CP}}$ using Equations~\eqref{eq:alpha_cp} and~\eqref{eq:alpha_cp2} until $C_{\text{CP}} \geq C$. We can prove the following facts to justify this process (see Zarei et al.~\cite{zarei:hscc:2020}). First, with probability $1$, the confidence level converges to $1$ in this process; thus, this process always terminates. Second, whenever this process stops, the statistical assertion has a confidence level of \emph{at~least}~$C$. 

Finally, an important adaptation of SMC for computer systems is presented by Mazurek et al.~\cite{mazurek:micro:2023}, which provides a confidence interval for the actual value of satisfaction probability \( p_\varphi \) in Equation~\eqref{eq:goal} via several SMC runs.
This can help evaluating the exact level of computer security in terms of the satisfaction probability of the security property $\varphi$. Specifically, this approach repeatedly runs SMC with different threshold values \( F \) on the same set of samples, determining the range of possible values for \( p_\varphi \) based on those SMC results.

\section{Usage Model}

\renewcommand{\arraystretch}{1.1}  

Our goal is to be able to ask---and rigorously answer---questions about security side channels.  To explain how we achieve this goal, we use three side channel case studies that we list below and dive into more deeply in three dedicated sections of this paper.

\begin{itemize}
    \item Prime \& Probe (P\&P) ~\cite{tromer:jcrypto:2010}: We explore this classic, well-studied side channel that leaks information through cache replacements.  
    \item ScatterCache~\cite{werner:css:2019} and Mirage~\cite{saileshwar:usenix-security:2021}: We quantitatively compare the security of these two techniques for defending a cache side channel.
    \item CleanupSpec~\cite{saileshwar:micro:2019} and unXpec~\cite{li:hpca:2022}: We quantitatively compare the security of these two techniques for defending against a Spectre-like attack and evaluate a new variant.
\end{itemize}

For each question we seek to answer with SMC, we develop an $experiment$ that consists of three steps.

\vspace{2pt}
\noindent
\textbf{Obtain Samples.}  The first step is running our simulator multiple times to obtain multiple data samples.  We explain in Section~\ref{sec:methodology} how we use our simulator to model probabilistic behaviors.  One key feature of SMC is that this step of obtaining data does \underline{not} require us to build any new probabilistic or formal models; we can simply use a typical cycle-by-cycle simulator with all of its detail.  Another key feature of SMC is that, for the type of properties we explore, SMC typically requires relatively few samples to achieve high levels of statistical confidence; most of our experiments require on the order of 20 samples to achieve 90-95\%~confidence.

\vspace{2pt}
\noindent
\textbf{Process Samples.}  For some experiments, we need to process the data we obtain.  For example,  
when asking a question about attack success versus observed noise level, we 
may discard data from executions whose noise levels were outside our range of interest.

\vspace{2pt}
\noindent
\textbf{Evaluate SMC Property Based on Processed Samples.}  As explained earlier, SMC evaluates binary properties, e.g., is attack accuracy greater than a given threshold?  These properties can be complicated expressions written in STL, but often simple properties suffice. 







\section{Simulation Methodology}
\label{sec:methodology}

Our technique can be applied to executions of real or simulated machines.  We focus here on simulation-based experimentation, because of its prevalence in the research community and its ability to study systems that have not been physically built.

We model multicore processors with gem5~\cite{lowe-power:gem5:2020}, which provides detailed, cycle-by-cycle, full-system simulation.  Because gem5, like most simulators, is deterministic, we must
inject any desired probabilistic behaviors. There are several sources of probabilistic behaviors, and we simulate them differently.\footnote{Real machines naturally exhibit probabilistic behaviors and thus would not require these behaviors to be injected during executions.} 

An important source of probabilistic behavior is natural (i.e., not introduced by the victim to obfuscate its behavior).  One such natural cause is the behavior of concurrently running processes.  
To study natural probabilistic behavior while still enabling repeatable results, we model this natural probabilistic behavior---which we will now refer to simply as \textit{natural noise}---by introducing a concurrent process that performs a simple loop.  In each loop iteration, the process waits some amount of time and then reads from a random memory address.\footnote{There is no true randomness in a computer program, but we use rand() to generate pseudo-random numbers.  By choosing the seed to rand(), we can repeat experiments.}  The wait time is a configurable random variable, and we consider 5 noise levels, in which a higher noise level corresponds to a shorter wait time. 
This natural noise model is chosen for its simplicity, but we could have chosen any other model; the choice of noise model is orthogonal to this work.

Another source of natural probabilistic behavior is randomness in cache replacement algorithms.  We can simulate caches with any replacement algorithm, some of which have probabilistic behaviors.  For example, NMRU (not most recently used), for associativity greater than 2, randomly chooses the victim from among the NMRU blocks in the set.

Finally, in some of our case studies, there are additional sources of probabilistic behavior---including intentional noise injected by a victim---and we discuss those sources then.

It is important to note that the natural noise model implemented in this work is chosen for its simplicity, but we could have chosen any other model. This work evaluates microarchitecture side channels with statistical rigor in the presence of noise, but the specific source or complexity of noise is orthogonal to this work. SMC is distribution agnostic, which means that the statistical guarantees on the observations still hold true regardless of the shape of the observation distribution. Whether the disturbance comes from a naïve pseudo-random loop, a full workload replay, hardware prefetchers, or deliberately injected variability, these conditions still apply because the source’s internal complexity does not enter into the Clopper–Pearson mathematics. Consequently, once we calibrate a noise generator to the rate of interference we wish to study, the subsequent probability estimates and confidence guarantees are fully transferable to any other generator that yields the same observable rate. In short, our evaluation methodology is deliberately agnostic to the provenance of noise and thus retains its statistical validity across a spectrum of realism and complexity.
\section{Case Study 1: Prime + Probe} \label{sec:case1}

We first illustrate the power and potential of SMC in the context of a simple, well-known side channel attack: prime+probe.  Our goal here is not to innovate in security, but rather to demonstrate the application of SMC to security.  The innovation lies in not just the use of SMC but also in the development of SMC experiments that provide insight and actionable information.

\subsection{Prime \& Probe in General}

In general, in a prime+probe (P\&P) attack, the attacker first $primes$ a processor's cache by placing its own data in the cache.  The attacker prompts the victim to run (e.g., by making a request to the victim to decode an input codeword).  In the course of doing its work, the victim brings its own data into the cache, thus displacing some of the data the attacker had placed there.  Then the attacker $probes$ the cache by reading the data it had originally put in the cache.  Any accesses that take longer than the latency of a cache hit indicate blocks that were evicted to make room for the victim's data.  The attacker can thus glean information about the victim by knowing which cache sets the victim accessed.  

There are many variations of the P\&P attack, but they all follow this basic structure.  For our purposes, many details of the attacks are orthogonal to what we are doing here, such as how the attacker determines which virtual addresses map to the same cache set after they have been translated.

\subsection{The Attack and Probabilistic Environment}

The victim program is performing AES decryption.  P\&P is applicable to many possible victim applications, but AES is a well-known example.  AES uses a 16-byte key that is the target of the attacker.  For each AES transmission, the program encrypts and sends 64 bytes of data across a channel. To deter attackers, we assume the program changes the encryption key every 2,500 transmissions. Thus an attacker has 2,500 P\&P iterations to inform its probabilistic guess of the key (which would allow them to decrypt all data sent with that key). The percent of time an attacker can break the key directly translates to the percent of data that can be decrypted by an attacker in the long run.

The attacker primes the cache, then prompts the victim to encrypt a codeword, and then probes the cache.  By performing this P\&P loop (i.e., prime, encrypt, probe) repeatedly, the attacker can determine which cache sets the victim is accessing when it accesses the AES lookup tables.  This cache access information enables the attacker to uncover the secret key at the half-byte granularity.  

In the absence of noise (either natural or injected), an attacker can obtain the secret key in relatively few P\&P iterations~\cite{tromer:jcrypto:2010}, which motivates the periodic changing of the key. However,
the introduction of noise makes the attacker's job more difficult.  For now, we consider only "natural" noise, as described in Section~\ref{sec:methodology}.  Later in this section, we consider injected noise.

Noise manifests as added memory accesses that are, to the attacker, indistinguishable from normal accesses made by the victim.  The "noise" accesses can cause the eviction of the attacker's data from the cache, which is then misinterpreted by the attacker as victim behavior.  More frequent noise accesses correspondingly cause more trouble for the attacker, i.e., the signal-to-noise ratio decreases. 

\subsection{Three Questions to Answer}  

We present three questions we would like to be able to answer about this attack.  We show how to use SMC to answer these questions that current methodologies cannot address.

\vspace{0.1cm}

\noindent
(Question 1.1) What is the probability that an attacker can successfully guess the key within 2,500 transmissions, as a function of the amount of noise in the system?  Specifically, we seek a specific level of statistical confidence that this probability is within a given interval.

\noindent
(Question 1.2) How much additional noise do we need to inject as a defense mechanism to achieve the same confidence that the attack success will be at some desired lower level?

\noindent
(Question 1.3) How is the probability of attack success affected by the cache replacement policy, with a focus on the impact of probabilistic behavior introduced when using LRU replacement?

\subsection{SMC Methodology} \label{subsec:smc-methodology}

In all P\&P experiments, we use SMC to study the probability of attack success, and we consider an attack to succeed only if it correctly guesses all 16 bytes of the key.  Because we are strictly focused on a probability, rather than a quantity (e.g., number of bytes leaked), we do not need the full power of SMC.\footnote{We will use full SMC in our latter two case studies.}  We can instead use SMC's statistical underpinnings without overlaying a temporal property on top.  We treat the experimental results (success/failure) as a Bernoulli random variable and use the Clopper-Pearson method to obtain confidence intervals.


\subsection{Results for Question 1.1}

\begin{figure}
    \centering
    \includegraphics[width=0.55\textwidth]{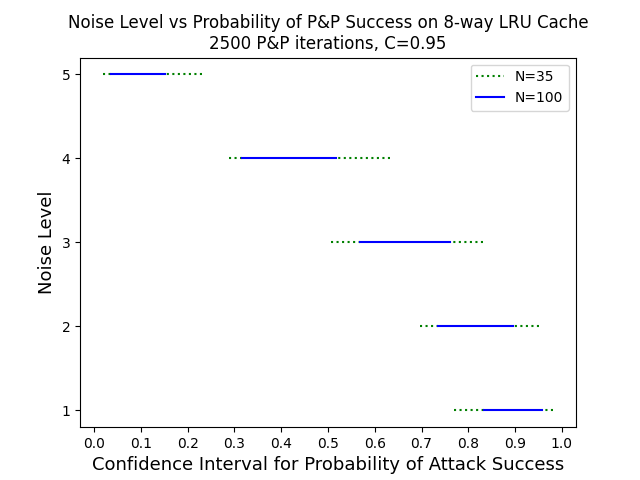}
    \caption{Experiment 1.1a: P\&P Success Probability vs. Noise Level. We use the Clopper-Pearson method to build confidence intervals for the probability of attack success for 5 distinct levels measuring the magnitude of injected noise. As one collects more data (i.e., for larger $N$), one can make more precise assertions about the security confidence interval.}
    \label{fig:case1:accuracy-vs-noise}
\end{figure}

\noindent
\textbf{Experiment 1.1a: At a given noise level, what is the probability that an attack will succeed?}
We fix the number of P\&P iterations at 2,500 and consider each of 5 noise levels. 
In Figure~\ref{fig:case1:accuracy-vs-noise}, we plot two 95\% confidence intervals for their corresponding attack success probabilities: one is built from 35 attacks (i.e., N=35) and one from 100. It is not surprising that greater noise leads to lesser accuracy, but the use of SMC enables us to rigorously quantify this relationship. For example, our intervals with N=35 tell us an attacker could successfully guess the key between 77\% and 98\% of the time at noise level 1, versus between only 2\% and 23\% at noise level 5. When we increase the number of attacks to N=100, we get tighter bounds of between 83\% and 96\% at noise level 1 and 3\% and 15\% for noise level 5.

\begin{figure}
    \centering
    \includegraphics[width=0.45\textwidth]{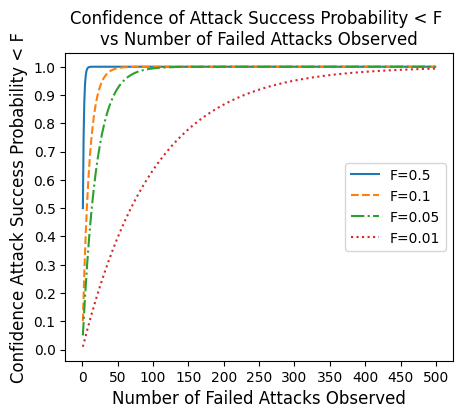}
    \caption{Confidence Attack Success Probability $<F$ vs Number of Failed Attacks Observed for different values of $F$. This experiment empirically shows that in experiments in which we only observe failed attacks, we must collect significantly more samples to be confident in high (e.g., >=99\%) security.}
    \label{fig:case1:confidence-vs-samples}
\end{figure}

When attack success is rare, we might have an entire set of $N$ samples in which the attack fails. For that situation, we further explore the relationship between the number of attacks and the tightness of our bounds.  Specifically, if we observe a number of attacks $N$ that have all failed, we would like to know how confident we can be that the attack will succeed with probability less than $F$, for different values of $F$. In Figure \ref{fig:case1:confidence-vs-samples}, we plot these results. As expected, to achieve a tighter bound on success probability requires more samples, but the question is how many.  For example, to reach 95\% confidence, it takes only 4 samples for $F$=0.5, but 28, 58, and 298 samples for $F$=0.1, 0.05, and 0.01, respectively.

\vspace{0.2cm}

\vspace{0.2cm}

\begin{figure}
    \centering
    \includegraphics[width=0.65\textwidth]{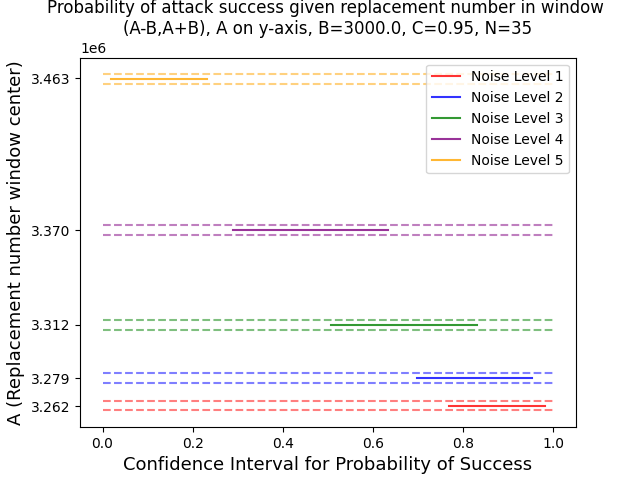}
    \caption{Experiment 1.1b: P\&P Success Probability vs. Number of Cache Replacements. 
    We evaluate the probability that an attack will succeed given a cache replacement number in the range (A-B, A+B). We use the Clopper-Pearson method to generate confidence intervals for the attack's success probability in the varying windows of the number of cache replacements for our 5 different noise levels. With this information, we can monitor system security with cache replacements as a proxy, and potentially inject noise to artificially drive up replacements to maintain a desired security level.}
    \label{fig:case1:accuracy-vs-replacements}
\end{figure}

\noindent
\textbf{Experiment 1.1b:  For a number of cache replacements in the range (A-B, A+B), what is the probability an attack will succeed?}  
The previous experiment provides insight, but it is not actionable; the victim is unlikely to be able to measure the noise level at runtime.  We now use cache replacements as a proxy for noise, with the advantage that they can be measured at runtime with hardware performance counters. In Figure~\ref{fig:case1:accuracy-vs-replacements}, we show the results for B=3,000 and 
 all five noise levels.  At each noise level we choose a value of A such that the number of replacements for each execution fall in the range (A-B, A+B). Each noise level corresponds to a color, and the range (A-B, A+B) for each noise level is denoted with a pair of horizontal dashed lines in the color for that noise level.   
 
We plot the confidence interval for accuracy for each value of A.
As expected, when there are more replacements---because there are more noise accesses leading to replacements)---it is more difficult for the attacker.  What is new is that we can rigorously quantify \textit{how much more difficult} it is for the attacker.  A victim can use this graph to drive an algorithm for monitoring replacements and either (a) simply knowing when it is vulnerable, or (b) injecting noise (discussed next) to drive up replacements when the number of replacements is less than desired.

\subsection{Results for Question 1.2}

While a victim could blindly inject noise to deter an attack, the results from SMC can inspire noise injection heuristics that limit how much noise needs to be injected (and thus limits the performance degradation due to injected noise).

From the previous experiments, a victim could determine how much noise to (continuously) inject to reach the noise level that corresponds to the desired level of attack success.  
In this section, we instead consider a victim that seeks to inject intermittent (rather than continuous) noise as a defense.

\begin{figure}
    \centering
    \includegraphics[width=0.9\textwidth]{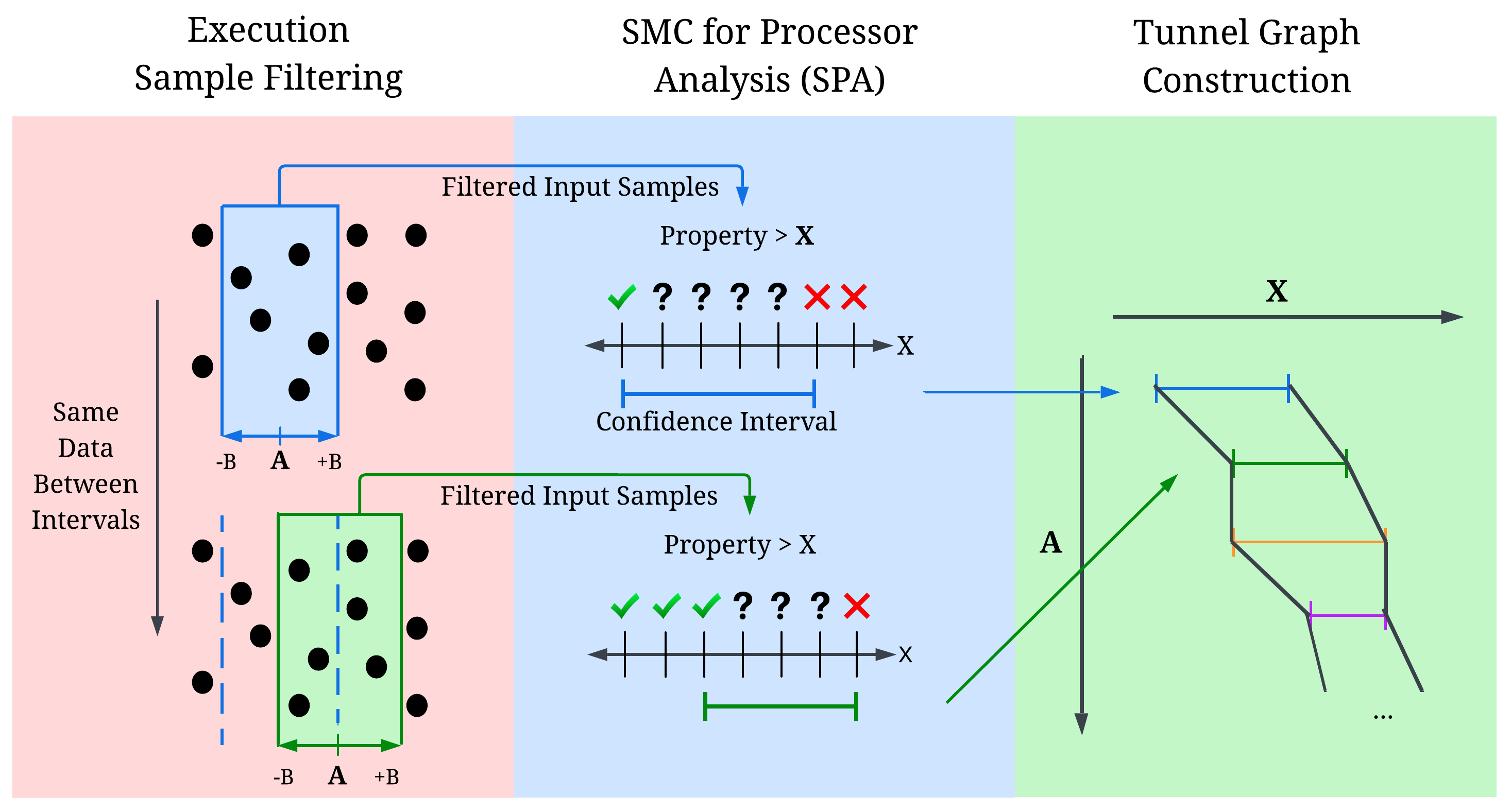}
    \caption{Tunnel Graph Creation Methodology. Similar to how SPA~\cite{mazurek:micro:2023} creates a confidence interval by iterating on a discrete metric in an SMC property along one dimension, a tunnel graph creates 2-dimensional intervals by: iterating on discrete input data filters, constructing new confidence intervals (either with SPA or Clopper-Pearson), and connecting their bounds in a 2D graph. Since each confidence interval adheres to a certain, user-specified confidence level, our entire tunnel graph does as well. Making the iterations in both dimensions more granular provides more continuous insights.}
    \label{fig:tunnel_graph_diagram}
\end{figure}

\vspace{0.2cm}

\noindent
\textbf{Experiment 1.2: If the victim injects noise in a fraction of loops that is in the range (A-B, A+B), is the attacker's accuracy greater than a given threshold?}  

Before presenting the results of this experiment, we first explain our methodology for presenting the results.  As described in Figure \ref{fig:tunnel_graph_diagram}, we construct a 2-dimensional confidence interval that we refer to as a \textit{tunnel graph}.  We iterate on a parameter value or a window of values along the x-axis (e.g., fraction of loops in which we inject noise). 
Using the data we sampled and separated along the x-axis, we iterate along the y-axis to construct accuracy confidence intervals using SMC with the same technique proposed by Mazurek et al.\cite{mazurek:micro:2023}. It is important to note that the tunnel graph is discrete in both dimensions at a specified granularity. A smaller granularity improves the continuity of the tunnel graph but increases the required number of simulations.

\begin{figure}
    \centering
    \includegraphics[width=0.55\textwidth]{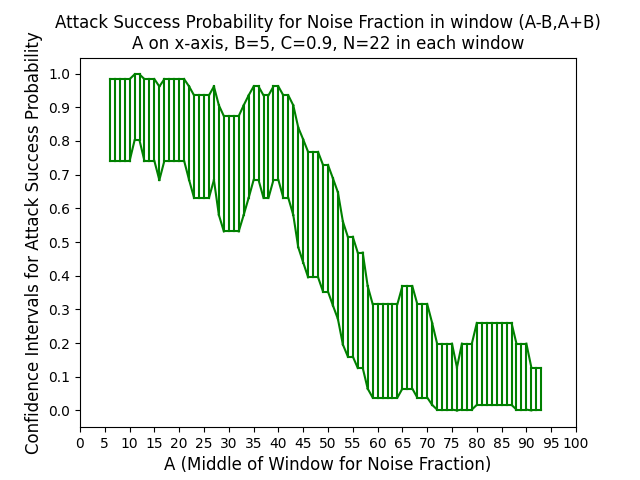}
    \caption{Experiment 1.2: P\&P Attack Success Probability vs. Frequency of Noise Injection. We create a tunnel graph to evaluate the relationship between obfuscation and success probability. As outlined in Figure \ref{fig:tunnel_graph_diagram}, we iterate on a window of noise magnitude to filter samples. We use the Clopper-Pearson method to generate 90\% confidence intervals on the attack success probability at each noise window and thus create a 90\% confidence tunnel. Noise does not manifest in discrete levels in real systems so it is valuable to have a rigorous \textit{and} continuous understanding of the relationship between noise and attack success.}
    \label{fig:case1:accuracy-vs-noise-frequency}
\end{figure}

Returning to Experiment 1.2, we fixed the noise level at 5 and decreased the number of attack iterations from our default of 2,500 down to 2,000.
In Figure~\ref{fig:case1:accuracy-vs-noise-frequency}, we plot the tunnel graph results for B=5 (a parameter that can be chosen by the victim), with A as the free variable on the x-axis. For example, when A=50, we are considering the situation in which the victim injects noise in 45-55\% of the loops.
We fix the confidence at $C$=90\%.  These results can be used to inform a defense in which noise is not injected all the time.  For example, if A=75 (i.e., we inject noise in 70-80\% of loops), the confidence interval indicates that only between 0.2\% and 20\% of attacks will succeed.  
These results are actionable---they tell the victim how often to inject noise---and lead to statistically rigorous guarantees on attack success.


\subsection{Results for Question 1.3}

Cache replacement policies can introduce a measure of probabilistic behavior, and we now explore the impact of this probabilistic behavior on the attacker's success.  



 \begin{figure}
    \centering
    \includegraphics[width=0.45\textwidth]{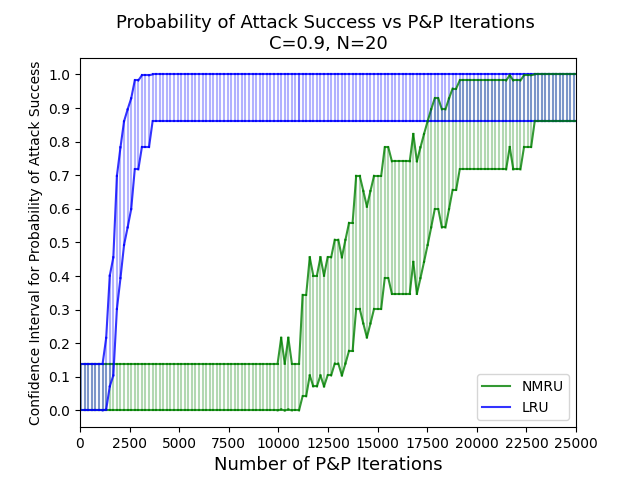}
    \caption{Experiment 1.3a: P\&P Iterations vs. Success Probability for LRU vs NMRU cache replacement. We create a tunnel graph showing the continuous relationship between the number of attack iterations and attack success for both eviction schemes. Readers can use these insights to make more informed decisions for their systems' caching schemes by balancing the performance advantages of LRU against the security advantages of NMRU.}
    \label{fig:c1-p4}
\end{figure}

 \vspace{0.2cm}

\noindent
\textbf{Experiment 1.3: At a fixed noise level, after X iterations, what is the probability of attack success?}
Now, we compare 8-way caches with LRU and NMRU replacement.  With NMRU, among the not-most-recently-used blocks in the set, the cache chooses the victim randomly.  This introduces probabilistic behavior that is more pronounced with greater associativity. (Consider the extreme case of a fully-associative cache, in which NMRU randomly chooses a victim from any block, except the LRU block, in the entire cache.) Figure~\ref{fig:c1-p4} shows that NMRU replacement has a significant impact on the success of an attack.  With LRU, an attacker needs only about 4,000 P\&P iterations to be 90\% confident of a successful attack; with NMRU, it takes nearly 24,000 P\&P iterations to achieve comparable results.  An architect can use this information when trading off the performance advantage of LRU against the security and implementation simplicity of NMRU.
The results in Figure~\ref{fig:c1-p4} also enable a victim to confidently choose how often to change the key for either choice of cache replacement algorithm.

\subsection{Conclusions from Case Study 1}

In this first case study, we have intentionally chosen 
the well-studied security issue of Prime and Probe.  In the context of P\&P, we have shown the power of SMC to provide greater statistical rigor and additional insights.  In doing so, we built no simplifying models and made no assumptions about probability distributions.
Our results here do not upend conventional wisdom, but they provide quantitative, actionable results that can be confidently used by victims.

\section{Case Study 2: ScatterCache \& Mirage}

In this case study, we compare two schemes for defending against cache side-channel attacks:
ScatterCache~\cite{werner:css:2019} and the subsequent Mirage~\cite{saileshwar:usenix-security:2021}.  Our goal is to highlight how SMC, combined with appropriate properties, can reveal insights when comparing security schemes.  

\subsection{Threat Model}

The attacker and victim are separate processes that share a cache (e.g., running on separate cores that share an L2 cache or last-level cache).  
The attacker seeks to determine accesses it can perform that can cause a victim's block to be evicted.  In a traditional set-associative cache, the problem devolves to the attacker seeking to identify accesses it can make that map to the same cache set accessed by the victim.  Hence these events were dubbed \textit{set-associative evictions (SAEs)}. In the non-traditional caches described next, there are no sets, but the term SAE is still used; it describes an eviction that was not caused by a capacity miss.  If an attacker's access to block X can cause a victim block Y to be evicted (and not due to capacity), X is considered part of Y's \textit{eviction set}.

\subsection{ScatterCache and Mirage}

Because it is relatively straightforward to find accesses that map to the same set in a traditional cache, there has been a substantial amount of security research to design caches that do not have traditional sets. ScatterCache cleverly borrows the idea of the skewed associative cache~\cite{seznec:isca:1993}, which breaks the basic idea of sets.  Each way of a skewed associative cache has a different mapping from block address to where that block can go in that way.  Thus, even if a block X maps to the 4th and 21st entries in ways 0 and 1, respectively, another block Y could map to the 53rd and 21st entries in ways 0 and 1.  This use of multiple mappings makes it more difficult for the attacker to identify blocks in the victim block's eviction set. Scatter-Cache's probabilistic behavior derives from the random seed it uses for changing the mappings.

Despite the added protection of ScatterCache, it was later shown to be vulnerable to Probabilistic Prime \& Probe (PPP)~\cite{purnal:sp:2021}, which can quickly construct eviction sets in only O(n) cache accesses.

A subsequent defense, Mirage, observed that ScatterCache's weakness was that any given eviction from a W-way skewed associative cache is choosing from among only the W locations to which the incoming block maps. Using a level of indirection, Mirage decouples mapping from replacement, enabling fully (pseudo-)random replacement. Random replacement, which is an inherently probabilistic behavior, greatly increases the difficulty of an attack. 

Our descriptions of ScatterCache and Mirage are brief and high-level because that is all that is required for understanding this case study.

\subsection{Question to Answer}

We want to compare the security of ScatterCache and Mirage 
and have statistical confidence in the result.  

\noindent
(Question 2) For each scheme, what is the number of observed SAEs while monitoring a fixed number of target addresses?

\subsection{SMC Methodology}

SMC is well-suited to comparing ScatterCache and Mirage because of the inherently probabilistic nature of the attacks and the Mirage scheme itself. By applying SMC, we want to quantify Mirage's advantage over ScatterCache. 

For our experiments, we use the gem5 model released with Mirage.  The system model, for both Mirage and ScatterCache, includes a 16-way 8MB L2 cache.   We compare ScatterCache and Mirage based on the attacker's ability to find a certain number of SAEs for particular victim addresses. 

We take 20 samples (i.e., N=20) for each cache design and use SMC to build a confidence interval for the number of SAEs that can be found. 
 


\subsection{Results for Question 2}

\begin{figure}
    \centering
    \includegraphics[width=0.45\textwidth]{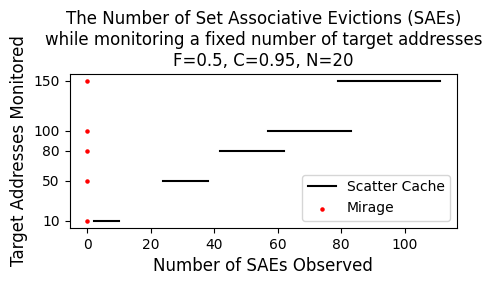}
    \caption{Experiment 2: Number of SAEs on target addresses vs. Number of target addresses. This experiment compares the security benefits of state-of-the-art probabilistic cache defense mechanisms. With appropriately chosen linear temporal logic properties, researchers can rigorously evaluate and compare attack schemes or defense mechanisms.}
    \label{fig:c2-p1}
\end{figure}

\noindent
\textbf{Experiment 2:  For a given number of cache accesses, is the number of SAEs on a group of target addresses greater than a threshold?}  In Figure \ref{fig:c2-p1} we show the results. Most strikingly, we observe no SAEs with Mirage, and SMC allows us to say with (at least) 95\% confidence that we will see zero SAEs.  Given the analytical model in the Mirage paper, this result is not surprising.  
The results for ScatterCache reveal a few phenomena.  First, ScatterCache is significantly less secure than Mirage.  Second, while the number of SAEs unsurprisingly increases as we track more target addresses, the increase is not perfectly linear.  Third, the confidence interval widens with the size of the group of target addresses.  

\subsection{Conclusions from Case Study 2}

 In Case Study 2, we have extended the application of SMC to evaluate two state-of-the-art cache defense mechanisms that are fundamentally probabilistic.  The results show that SMC can both confirm existing intuition (Mirage is more secure than ScatterCache) and rigorously quantify that difference in security.  This experiment provides architects with quantitative data that they can use to weigh the security benefits of Mirage against its added complexity (with respect to ScatterCache).
\section{Case Study 3: CleanupSpec / unXpec} \label{sec:case3}

In addition to cache side channels, which we have explored in the first two case studies, there also exist side channels in speculative processor cores.  Modern cores frequently speculate---on a branch outcome, whether a load will hit in the cache, etc.---and speculation can affect microarchitectural state in a way that can leak information.  The infamous Spectre vulnerability exploits such speculative microarchitectural state.  

\subsection{Defense: CleanupSpec}

Much recent work has sought to overcome this vulnerability by effectively hiding speculative changes to the microarchitectural state.  One representative scheme is CleanupSpec~\cite{saileshwar:micro:2019}, which "cleans up" (i.e., undoes) the effects of speculative execution on the cache.  To do this, CleanupSpec tracks any blocks evicted from the cache during speculative execution; if misspeculation is detected, it fetches those blocks back into the cache, replacing the blocks that were brought into the cache during speculative execution.

\subsection{Attack: unXpec}

Although cleaning up the speculative state seems sufficient, unXpec~\cite{li:hpca:2022} shows that information still leaks.  Specifically, the length of time needed to clean up the speculative state is a side channel, because it depends on the number of blocks that need to be evicted from the cache and the corresponding number of blocks that must be brought back to the cache. If the value of the secret data affects the number of blocks that must be evicted/filled, it creates a latency difference that leaks data. Experiments in the unXpec paper showed, in one example, that the latency difference between the secret bit being 0 and 1 was 32 cycles.

\subsection{Defense: Obfuscated Clean-Up Latency}



The side channel revealed by unXpec is the latency of the clean-up process.  An effective defense mechanism can obfuscate that channel by enforcing a constant-time rollback for undo schemes. That is, whenever a mis-speculation happens, the CPU core must be stalled for a constant time, even if the "clean up" takes less time to finish or needs no "clean up" at all. But this simple and intuitive countermeasure brings a significant performance overhead.  As reported in unXpec~\cite{li:hpca:2022}, the performance overhead for the SPECCPU 2017 benchmarks~\cite{bucek:icpe:2018} ranges from 22.4\% to 72.8\% with constant rollback time ranging from 25 to 65 cycles.

To mitigate this performance impact, we have explored the use of probabilistic obfuscation. Instead of extending the duration of every clean up, the processor extends it with a given probability that we refer to as the obfuscation probability. To the best of our knowledge, this obfuscation scheme has not been explored, but we do not claim that as the contribution of this section.  Rather, our goal is how to evaluate the effectiveness and performance impact of an unexplored and inherently probabilistic scheme.  

Processor cores---like the out-of-order core that we consider in this case study---are complicated systems that are difficult to accurately model with abstractions or simplifications.  Architects generally rely on cycle-by-cycle simulators because that level of detail is necessary to capture all of the subtle behaviors that can significantly impact performance.  Thus, to evaluate this probabilistic obfuscation scheme, we need SMC's features: the ability to use a highly detailed model and incorporate probabilistic behaviors.




\subsection{Question to Answer}

We seek to understand how much obfuscation is required to avoid side-channel attacks on the clean--up latency.

\noindent (Question 3) What obfuscation probability should the victim use to guarantee a sufficiently low probability of attack success?


\subsection{Results for Question 3}

In this experiment, we randomly generate a 1,000-bit secret and cause a mis-speculation while processing each bit of the secret.  If the secret bit is 0, no blocks need to be cleaned up; if the secret bit is 1, one block must be brought back into the cache.  We assume the attacker can measure the time taken to recover from mis-speculation and use that information to infer the value of a secret bit.  We define the attack accuracy as the fraction of correct secret bits the attacker obtains in this fashion.

\begin{figure}
    \centering
    \includegraphics[width=0.55\textwidth]{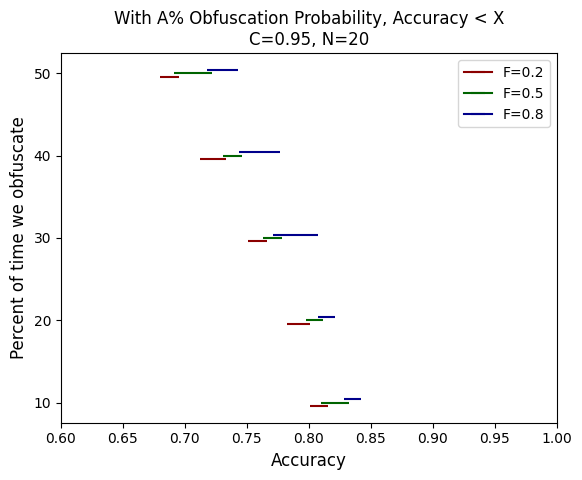}
    \caption{Experiment 3a: Obfuscation probability vs. accuracy. These results enable obfuscation defense schemes to quantitatively determine the resulting attack success probability.}
    \label{fig:c3-p1}
\end{figure}

\vspace{0.2cm}

\noindent
\textbf{Experiment 3a: At a given obfuscation probability, is the attacker's accuracy less than a given threshold?} In Figure~\ref{fig:c3-p1}, for each of the 3 SMC proportions ($F$), we show 95\% confidence intervals for 5 different obfuscation probabilities from 10\% to 50\%. (X\% means for every detected mis-speculation, the core has a X\% probability of adding a latency that can eliminate the timing difference.) While a higher obfuscation probability makes the performance worse (discussed at the end of Experiment 3b), these results can be used to understand the trade-off between obfuscation probability and attack accuracy.  For example, a defender could obfuscate only 50\% of the time and have 95\% confidence at proportion $F$=0.5 (i.e., the median) that the attacker's accuracy will be only between 69-72\%.  However, at 10\% obfuscation, 95\% confidence, and $F$=0.5, the attacker's accuracy will be between 81-83\%.

\vspace{0.2cm}
\begin{figure}
    \centering
    \includegraphics[width=0.55\textwidth]{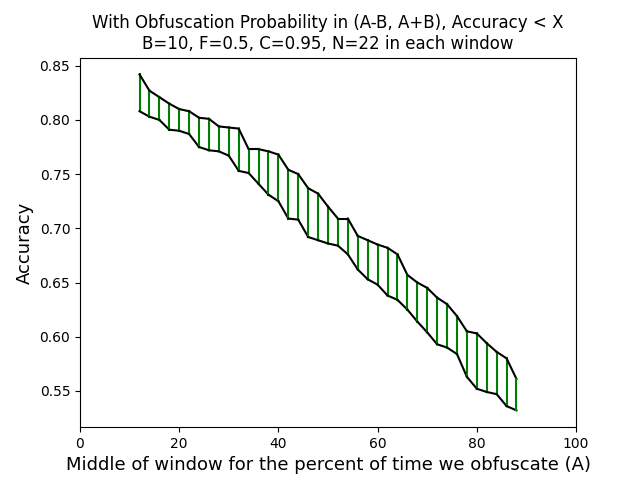}
    \caption{Experiment 3b: Obfuscation probability range vs. accuracy. We create a tunnel graph to evaluate the relationship between the magnitude of obfuscation and attack accuracy. As we increase the granularity in both dimensions, we observe near linear returns in side-channel security as we increase the amount that we obfuscate. Note that obfuscation has well-known performance overheads and therefore these results allow the defender to make a well-informed trade-off.}
    \label{fig:c3-p2}
\end{figure}

\noindent
\textbf{Experiment 3b: With obfuscation probability in the range (A-B, A+B) percent, is the attacker's accuracy less than a given threshold?} Whereas Experiment 3a gives confidence intervals for discrete obfuscation probabilities, this experiment allows us to consider ranges of obfuscation probabilities. In Figure \ref{fig:c3-p2}, we plot results for B=10 with A as a free variable on the x-axis. Results show that, for example, if the obfuscation probability is between 60\%-80\%, we have a 95\% confidence that accuracy will be less than 65\%.


The performance impact of probabilistic obfuscation is directly related to two factors: how often the benchmark misspeculates and the probability of obfuscation.  We simulated a range of obfuscation probabilities on a subset of the SPECCPU 2017 benchmarks that unXpec~\cite{li:hpca:2022} identified as being at the two extremes for performance impact.  Our results show that the probability of obfuscation has a linear impact on performance; 50\% obfuscation probability has 50\% the performance impact of 100\% obfuscation.  There is negligible variability in performance, despite the probabilistic behavior, because the performance overheads for a single-threaded application effectively just sum.  For example, at an obfuscation probability of 50\%, the performance will be almost identical, regardless of which 50\% of the misspeculations are chosen for obfuscation.  As such, we do not present this as an SMC Experiment or graph.

\subsection{Conclusions from Case Study 3}

This third case study shows how we can apply SMC to explore a new probabilistic scheme for defending against speculative execution attacks.  Critically, SMC eliminates any need to build an abstract or simplified model of the processor under study, and it provides actionable insight into the trade-off between security and performance.

\section{Discussion}

We have demonstrated that SMC is a powerful tool for evaluating security, under the assumptions we outlined in Section~\ref{sec:assumptions}.  We now discuss two issues that computer architects and systems designers should be aware of when using SMC for security analysis.

\vspace{0.2cm}
\noindent{\bf{Real Systems. }} 
SMC evaluation applies to both the simulated systems we focus on in this paper, as well as real hardware.  We focus on simulation because we often want to study systems that do not exist (yet).  SMC is a great match for simulation, in that SMC requires relatively few samples to achieve statistical confidence; because simulation is often time-consuming, it is a major benefit to minimize the number of required simulations.  But SMC can also be of great benefit when studying real hardware, particularly if those experiments are long. 

\vspace{0.2cm}
\noindent{\bf{Injecting Variability. }}
If we are experimenting using simulation, we may find ourselves needing to inject variability, because simulators are usually deterministic.  Depending on the source of variability that we wish to model, we need to devise a variability injection scheme that mimics the variability we would expect to see in real hardware.  We leave a study of how best to inject variability to future work.  It is important to note that our use of SMC is orthogonal to the methodology used for injecting variability.
\section{Related Work in Security Evaluation}

Many papers have been published on formal verification and evaluation for computer security, and we focus here on those papers that are most closely related to our work.  

\subsection{Security Verification with Formal Methods}

Formal methods, which employ abstract symbolic mathematics to formulate and algorithmically solve complex system engineering problems, are a powerful tool for the security analysis of computer systems. They first emerge as high-level verification tools capable of answering whether certain types of attacks are possible. Specifically, system executions can be abstracted using symbolic models, and formal verification algorithms can then be applied to determine whether the model is completely secure or if vulnerabilities exist. For example, Deng et al. have applied formal methods to evaluate cache side-channel attacks~\cite{deng:hasp:2018,deng:jhss:2019,deng:ieeetc:2021}.

However, it is important to note that the guarantees provided by formal verification are only as strong as the underlying models. These guarantees are valid only if all possible model executions are adequately represented in the abstract model. If the model is overly abstract and fails to capture factors that contribute to security issues, it must be refined to include those factors. Side channels, as an example, were only recently included to allow formal analysis~\cite{yang:isca:2023}.

While using detailed model in formal verification improves the comprehensiveness of security guarantees, it also introduces challenges in computational feasibility. This is true not only in terms of the cost of constructing the models, but more importantly in the complexity of solving them using formal methods. Broadly, there are two main approaches to verifying abstract symbolic models. The first is model checking, which exhaustively explores all possible model paths to determine whether a security property holds. The second is theorem proving, which aims to derive a formal proof that the model satisfies the desired security property from the rules of the model. Both approaches experience significant growth in computational cost as the size and complexity of the model increase.

\subsection{Evaluation with Probabilistic Models}

Probabilism is a critical consideration in evaluating system security as defense mechanisms typically rely on randomization (e.g., injecting obfuscation noise) to mitigate attacks. Quantitatively assessing the probability of successful attacks/defenses is essential for understanding the actual level of security in these cases. 

One way to handle this issue is to develop probabilistic models that capture the behaviors relevant to security analysis. 
Such models generally involve characterizing behaviors (e.g., whether a memory access is to secret data) with fixed probabilities or probability distributions.  Many probabilistic models have been developed, including for the study of cache side channels~\cite{deutsch:isca:2023,domnitser:cns:2010,bourgeat:micro:2020, genkin:asiaccs:2023,he:micro:2017}.

Probabilistic models, while a natural fit, often have significant shortcomings.  They often involve making assumptions about distributions (e.g., a distribution is Gaussian) so that the models can be solved analytically; the confidence in the results of such models are thus tied to the confidence in these assumptions.  
Thus, the results derived from these probabilistic models (even when using formal methods) are contingent upon the validity of the underlying assumptions and simplifications.

SMC does not require constructing explicit probabilistic models; instead, it can directly and rigorously analyze execution samples against various security properties of interest. SMC can also be applied to existing probabilistic models that are still too complex (despite assumptions and simplifications) to have closed-form solutions and thus rely on statistical solutions like Monte Carlo~\cite{deutsch:isca:2023}.
While SMC provides high-confidence rather than absolute guarantees, this level of assurance is usually sufficient for most practical applications.

\subsection{Security Metrics}

In any evaluation, security or otherwise, the choice of metrics is critical.  While the choices of metrics for performance are generally straightforward, metrics for security are not always as clear-cut.  As such, researchers have proposed new security metrics that provide additional insights.  Two notable examples are the side-channel vulnerability factor (SVF)~\cite{demme:isca:2012} and subsequent cache side-channel vulnerability factor (CSCV)~\cite{zhang:hasp:2013}.  

The development of new metrics is complementary to our work. Moreover, SMC offers a unified method for rigorously evaluating the system against these security metrics.
For example, with SMC we can determine with statistical confidence whether a metric like SVF or CSV will be less than a desired threshold.

\subsection{Evaluation with SMC}

Mazurek et al.~\cite{mazurek:micro:2023} took the established SMC methodology and showed how to apply it in the context of computer architecture.  To our knowledge, it was the first work to do so, and it additionally introduced an SMC extension that enabled them to produce confidence intervals for metric values (e.g., runtime).  The paper is limited, though, to only the simplest of SMC properties---is a given metric greater/less than a threshold---and it does not provide any experiments that would be insightful for evaluating security.  Rather, it focused on issues of performance and event counting (e.g., cache misses).
\section{Conclusions}

Probabilistic behaviors are fundamental to many side channels, and they complicate the evaluation of security attacks and defenses. We propose using statistical model checking to address this challenge.  SMC allows an architect to evaluate expressive properties with statistical rigor, without the use of abstract or simplified system models.  The architect can use a cycle-by-cycle simulator, which provides greater system detail, and does not need to spend time developing a probabilistic or formal model.  

Through three case studies, we hope to persuade computer architects and, more broadly, system designers to adopt SMC---using experiments we have developed, plus others they might develop---and thus improve the quality of security evaluations.  Potential future directions include the development of new temporal logic properties that provide insight into security and the design and evaluation of variability injection methodologies.

\section{Acknowledgments}
This work is supported in part by the National Science Foundation under grant CCF-213-3160.

\bibliographystyle{ACM-Reference-Format}
\bibliography{refs,security,shepherding-refs,yu}


\begin{thebibliography}{33}


\ifx \showCODEN    \undefined \def \showCODEN     #1{\unskip}     \fi
\ifx \showISBNx    \undefined \def \showISBNx     #1{\unskip}     \fi
\ifx \showISBNxiii \undefined \def \showISBNxiii  #1{\unskip}     \fi
\ifx \showISSN     \undefined \def \showISSN      #1{\unskip}     \fi
\ifx \showLCCN     \undefined \def \showLCCN      #1{\unskip}     \fi
\ifx \shownote     \undefined \def \shownote      #1{#1}          \fi
\ifx \showarticletitle \undefined \def \showarticletitle #1{#1}   \fi
\ifx \showURL      \undefined \def \showURL       {\relax}        \fi
\providecommand\bibfield[2]{#2}
\providecommand\bibinfo[2]{#2}
\providecommand\natexlab[1]{#1}
\providecommand\showeprint[2][]{arXiv:#2}

\bibitem[Agha and Palmskog(2018)]%
        {agha2018survey}
\bibfield{author}{\bibinfo{person}{Gul Agha} {and} \bibinfo{person}{Karl Palmskog}.} \bibinfo{year}{2018}\natexlab{}.
\newblock \showarticletitle{A survey of statistical model checking}.
\newblock \bibinfo{journal}{\emph{ACM Transactions on Modeling and Computer Simulation}} \bibinfo{volume}{28}, \bibinfo{number}{1} (\bibinfo{year}{2018}), \bibinfo{pages}{6:1--6:39}.
\newblock


\bibitem[Alameldeen and Wood(2003)]%
        {alameldeen:hpca:2003}
\bibfield{author}{\bibinfo{person}{Alaa~R. Alameldeen} {and} \bibinfo{person}{David~A. Wood}.} \bibinfo{year}{2003}\natexlab{}.
\newblock \showarticletitle{Variability in Architectural Simulations of Multi-Threaded Workloads}. In \bibinfo{booktitle}{\emph{Proceedings of the 9th International Symposium on High-Performance Computer Architecture}}.
\newblock


\bibitem[Bourgeat et~al\mbox{.}(2020)]%
        {bourgeat:micro:2020}
\bibfield{author}{\bibinfo{person}{Thomas Bourgeat}, \bibinfo{person}{Jules Drean}, \bibinfo{person}{Yuheng Yang}, \bibinfo{person}{Lillian Tsai}, \bibinfo{person}{Joel Emer}, {and} \bibinfo{person}{Mengjia Yan}.} \bibinfo{year}{2020}\natexlab{}.
\newblock \showarticletitle{CaSA: End-to-end Quantitative Security Analysis of Randomly Mapped Caches}. In \bibinfo{booktitle}{\emph{2020 53rd Annual IEEE/ACM International Symposium on Microarchitecture (MICRO)}}. \bibinfo{pages}{1110--1123}.
\newblock


\bibitem[Bucek et~al\mbox{.}(2018)]%
        {bucek:icpe:2018}
\bibfield{author}{\bibinfo{person}{James Bucek}, \bibinfo{person}{Klaus-Dieter Lange}, {and} \bibinfo{person}{J\'{o}akim v. Kistowski}.} \bibinfo{year}{2018}\natexlab{}.
\newblock \showarticletitle{SPEC CPU2017: Next-Generation Compute Benchmark}. In \bibinfo{booktitle}{\emph{Companion of the 2018 ACM/SPEC International Conference on Performance Engineering}}. \bibinfo{pages}{41–42}.
\newblock


\bibitem[Carlson et~al\mbox{.}(2011)]%
        {carlson:sc:2011}
\bibfield{author}{\bibinfo{person}{T.~E. Carlson}, \bibinfo{person}{W. Heirman}, {and} \bibinfo{person}{L. Eeckhout}.} \bibinfo{year}{2011}\natexlab{}.
\newblock \showarticletitle{Sniper: Exploring the Level of Abstraction for Scalable and Accurate Parallel Multi-Core Simulation}. In \bibinfo{booktitle}{\emph{SC}}.
\newblock


\bibitem[Chen et~al\mbox{.}(2012)]%
        {chen:hpca:2012}
\bibfield{author}{\bibinfo{person}{Tianshi Chen}, \bibinfo{person}{Yunji Chen}, \bibinfo{person}{Qi Guo}, \bibinfo{person}{Olivier Temam}, \bibinfo{person}{Yue Wu}, {and} \bibinfo{person}{Weiwu Hu}.} \bibinfo{year}{2012}\natexlab{}.
\newblock \showarticletitle{Statistical performance comparisons of computers}. In \bibinfo{booktitle}{\emph{IEEE International Symposium on High-Performance Comp Architecture}}.
\newblock


\bibitem[Clopper and Pearson(1934)]%
        {clopper1934use}
\bibfield{author}{\bibinfo{person}{Charles~J Clopper} {and} \bibinfo{person}{Egon~S Pearson}.} \bibinfo{year}{1934}\natexlab{}.
\newblock \showarticletitle{The use of confidence or fiducial limits illustrated in the case of the binomial}.
\newblock \bibinfo{journal}{\emph{Biometrika}} \bibinfo{volume}{26}, \bibinfo{number}{4} (\bibinfo{year}{1934}), \bibinfo{pages}{404--413}.
\newblock


\bibitem[Demme et~al\mbox{.}(2012)]%
        {demme:isca:2012}
\bibfield{author}{\bibinfo{person}{John Demme}, \bibinfo{person}{Robert Martin}, \bibinfo{person}{Adam Waksman}, {and} \bibinfo{person}{Simha Sethumadhavan}.} \bibinfo{year}{2012}\natexlab{}.
\newblock \showarticletitle{Side-channel Vulnerability Factor: A Metric for Measuring Information Leakage}. In \bibinfo{booktitle}{\emph{Proceedings of the 39th International Symposium on Computer Architecture}}.
\newblock


\bibitem[Deng et~al\mbox{.}(2021)]%
        {deng:ieeetc:2021}
\bibfield{author}{\bibinfo{person}{Shuwen Deng}, \bibinfo{person}{Nikolay Matyunin}, \bibinfo{person}{Wenjie Xiong}, \bibinfo{person}{Stefan Katzenbeisser}, {and} \bibinfo{person}{Jakub Szefer}.} \bibinfo{year}{2021}\natexlab{}.
\newblock \showarticletitle{Evaluation of cache attacks on arm processors and secure caches}.
\newblock \bibinfo{journal}{\emph{IEEE Trans. Comput.}} \bibinfo{volume}{71}, \bibinfo{number}{9} (\bibinfo{year}{2021}), \bibinfo{pages}{2248--2262}.
\newblock


\bibitem[Deng et~al\mbox{.}(2018)]%
        {deng:hasp:2018}
\bibfield{author}{\bibinfo{person}{Shuwen Deng}, \bibinfo{person}{Wenjie Xiong}, {and} \bibinfo{person}{Jakub Szefer}.} \bibinfo{year}{2018}\natexlab{}.
\newblock \showarticletitle{Cache timing side-channel vulnerability checking with computation tree logic}. In \bibinfo{booktitle}{\emph{Proceedings of the 7th International Workshop on Hardware and Architectural Support for Security and Privacy}}. \bibinfo{publisher}{Association for Computing Machinery}, \bibinfo{address}{New York, NY, USA}.
\newblock
\showISBNx{9781450365000}


\bibitem[Deng et~al\mbox{.}(2019)]%
        {deng:jhss:2019}
\bibfield{author}{\bibinfo{person}{Shuwen Deng}, \bibinfo{person}{Wenjie Xiong}, {and} \bibinfo{person}{Jakub Szefer}.} \bibinfo{year}{2019}\natexlab{}.
\newblock \showarticletitle{Analysis of Secure Caches Using a Three-Step Model for Timing-Based Attacks}.
\newblock \bibinfo{journal}{\emph{Journal of Hardware and Systems Security}}  \bibinfo{volume}{3} (\bibinfo{year}{2019}), \bibinfo{pages}{397--425}.
\newblock


\bibitem[Deutsch et~al\mbox{.}(2023)]%
        {deutsch:isca:2023}
\bibfield{author}{\bibinfo{person}{Peter~W. Deutsch}, \bibinfo{person}{Weon~Taek Na}, \bibinfo{person}{Thomas Bourgeat}, \bibinfo{person}{Joel~S. Emer}, {and} \bibinfo{person}{Mengjia Yan}.} \bibinfo{year}{2023}\natexlab{}.
\newblock \showarticletitle{Metior: A Comprehensive Model to Evaluate Obfuscating Side-Channel Defense Schemes}. In \bibinfo{booktitle}{\emph{Proceedings of the 50th Annual International Symposium on Computer Architecture}}.
\newblock


\bibitem[Domnitser et~al\mbox{.}(2010)]%
        {domnitser:cns:2010}
\bibfield{author}{\bibinfo{person}{Leonid Domnitser}, \bibinfo{person}{Nael Abu-Ghazaleh}, {and} \bibinfo{person}{Dmitry Ponomarev}.} \bibinfo{year}{2010}\natexlab{}.
\newblock \showarticletitle{A Predictive Model for Cache-Based Side Channels in Multicore and Multithreaded Microprocessors}. In \bibinfo{booktitle}{\emph{Computer Network Security}}.
\newblock


\bibitem[Genkin et~al\mbox{.}(2023)]%
        {genkin:asiaccs:2023}
\bibfield{author}{\bibinfo{person}{Daniel Genkin}, \bibinfo{person}{William Kosasih}, \bibinfo{person}{Fangfei Liu}, \bibinfo{person}{Anna Trikalinou}, \bibinfo{person}{Thomas Unterluggauer}, {and} \bibinfo{person}{Yuval Yarom}.} \bibinfo{year}{2023}\natexlab{}.
\newblock \showarticletitle{CacheFX: A Framework for Evaluating Cache Security}. In \bibinfo{booktitle}{\emph{Proceedings of the 2023 ACM Asia Conference on Computer and Communications Security}}. \bibinfo{pages}{163–176}.
\newblock


\bibitem[He and Lee(2017)]%
        {he:micro:2017}
\bibfield{author}{\bibinfo{person}{Zecheng He} {and} \bibinfo{person}{Ruby~B. Lee}.} \bibinfo{year}{2017}\natexlab{}.
\newblock \showarticletitle{How secure is your cache against side-channel attacks?}. In \bibinfo{booktitle}{\emph{Proceedings of the 50th Annual IEEE/ACM International Symposium on Microarchitecture}}. \bibinfo{pages}{341–353}.
\newblock


\bibitem[Irving et~al\mbox{.}(2020)]%
        {irving:fcs:2020}
\bibfield{author}{\bibinfo{person}{Samuel Irving}, \bibinfo{person}{Bin Li}, \bibinfo{person}{Shaoming Chen}, \bibinfo{person}{Lu Peng}, \bibinfo{person}{Weihua Zhang}, {and} \bibinfo{person}{Lide Duan}.} \bibinfo{year}{2020}\natexlab{}.
\newblock \showarticletitle{Computer Comparisons in the Presence of Performance Variation}.
\newblock \bibinfo{journal}{\emph{Frontiers of Computer Science}} \bibinfo{volume}{14}, \bibinfo{number}{1} (\bibinfo{year}{2020}).
\newblock


\bibitem[Kalibera and Jones(2020)]%
        {kalibera:arxiv:2007}
\bibfield{author}{\bibinfo{person}{Tomas Kalibera} {and} \bibinfo{person}{Richard Jones}.} \bibinfo{year}{2020}\natexlab{}.
\newblock \bibinfo{title}{Quantifying Performance Changes with Effect Size Confidence Intervals}.
\newblock
\showeprint[arxiv]{2007.10899}~[stat.ME]


\bibitem[Legay et~al\mbox{.}(2010)]%
        {legay2010statistical}
\bibfield{author}{\bibinfo{person}{A. Legay}, \bibinfo{person}{B. Delahaye}, {and} \bibinfo{person}{S. Bensalem}.} \bibinfo{year}{2010}\natexlab{}.
\newblock \showarticletitle{Statistical model checking: An overview}.
\newblock In \bibinfo{booktitle}{\emph{Runtime Verification}}, \bibfield{editor}{\bibinfo{person}{Howard Barringer}, \bibinfo{person}{Ylies Falcone}, \bibinfo{person}{Bernd Finkbeiner}, \bibinfo{person}{Klaus Havelund}, \bibinfo{person}{Insup Lee}, \bibinfo{person}{Gordon Pace}, \bibinfo{person}{Grigore Ro{\c s}u}, \bibinfo{person}{Oleg Sokolsky}, {and} \bibinfo{person}{Nikolai Tillmann}} (Eds.). Vol.~\bibinfo{volume}{6418}. \bibinfo{publisher}{Springer Berlin Heidelberg}.
\newblock


\bibitem[Li et~al\mbox{.}(2022)]%
        {li:hpca:2022}
\bibfield{author}{\bibinfo{person}{Mengming Li}, \bibinfo{person}{Chenlu Miao}, \bibinfo{person}{Yilong Yang}, {and} \bibinfo{person}{Kai Bu}.} \bibinfo{year}{2022}\natexlab{}.
\newblock \showarticletitle{unXpec: Breaking Undo-based Safe Speculation}. In \bibinfo{booktitle}{\emph{2022 IEEE International Symposium on High-Performance Computer Architecture (HPCA)}}. \bibinfo{pages}{98--112}.
\newblock


\bibitem[Lowe-Power et~al\mbox{.}(2020)]%
        {lowe-power:gem5:2020}
\bibfield{author}{\bibinfo{person}{J. Lowe-Power}, \bibinfo{person}{A.~M. Ahmad}, \bibinfo{person}{A. Akram}, \bibinfo{person}{M. Alian}, \bibinfo{person}{R. Amslinger}, \bibinfo{person}{M. Andreozzi}, \bibinfo{person}{A. Armejach}, \bibinfo{person}{N. Asmussen}, \bibinfo{person}{B. Beckmann}, \bibinfo{person}{S. Bharadwaj}, \bibinfo{person}{G. Black}, \bibinfo{person}{G. Bloom}, \bibinfo{person}{B.~R. Bruce}, \bibinfo{person}{D.~Rodrigues Carvalho}, \bibinfo{person}{J. Castrillon}, \bibinfo{person}{L. Chen}, \bibinfo{person}{N. Derumigny}, \bibinfo{person}{S. Diestelhorst}, \bibinfo{person}{W. Elsasser}, \bibinfo{person}{C. Escuin}, \bibinfo{person}{M. Fariborz}, \bibinfo{person}{A. Farmahini-Farahani}, \bibinfo{person}{P. Fotouhi}, \bibinfo{person}{R. Gambord}, \bibinfo{person}{J. Gandhi}, \bibinfo{person}{D. Gope}, \bibinfo{person}{T. Grass}, \bibinfo{person}{A. Gutierrez}, \bibinfo{person}{B. Hanindhito}, \bibinfo{person}{A. Hansson}, \bibinfo{person}{S. Haria}, \bibinfo{person}{A. Harris},
  \bibinfo{person}{T. Hayes}, \bibinfo{person}{A. Herrera}, \bibinfo{person}{M. Horsnell}, \bibinfo{person}{S.~Ali~Raza Jafri}, \bibinfo{person}{R. Jagtap}, \bibinfo{person}{H. Jang}, \bibinfo{person}{R. Jeyapaul}, \bibinfo{person}{T.~M. Jones}, \bibinfo{person}{M. Jung}, \bibinfo{person}{S. Kannoth}, \bibinfo{person}{H. Khaleghzadeh}, \bibinfo{person}{Y. Kodama}, \bibinfo{person}{T. Krishna}, \bibinfo{person}{T. Marinelli}, \bibinfo{person}{C. Menard}, \bibinfo{person}{A. Mondelli}, \bibinfo{person}{M. Moreto}, \bibinfo{person}{T. Mück}, \bibinfo{person}{O. Naji}, \bibinfo{person}{K. Nathella}, \bibinfo{person}{H. Nguyen}, \bibinfo{person}{N. Nikoleris}, \bibinfo{person}{L.~E. Olson}, \bibinfo{person}{M. Orr}, \bibinfo{person}{B. Pham}, \bibinfo{person}{P. Prieto}, \bibinfo{person}{T. Reddy}, \bibinfo{person}{A. Roelke}, \bibinfo{person}{M. Samani}, \bibinfo{person}{A. Sandberg}, \bibinfo{person}{J. Setoain}, \bibinfo{person}{B. Shingarov}, \bibinfo{person}{M.~D. Sinclair}, \bibinfo{person}{T. Ta},
  \bibinfo{person}{R. Thakur}, \bibinfo{person}{G. Travaglini}, \bibinfo{person}{M. Upton}, \bibinfo{person}{N. Vaish}, \bibinfo{person}{I. Vougioukas}, \bibinfo{person}{W. Wang}, \bibinfo{person}{Z. Wang}, \bibinfo{person}{N. Wehn}, \bibinfo{person}{C. Weis}, \bibinfo{person}{D.~A. Wood}, \bibinfo{person}{H. Yoon}, {and} \bibinfo{person}{E.~F. Zulian}.} \bibinfo{year}{2020}\natexlab{}.
\newblock \bibinfo{title}{The gem5 Simulator: Version 20.0+}.
\newblock
\showeprint[arxiv]{2007.03152}~[cs.AR]


\bibitem[Maler and Nickovic(2004)]%
        {maler2004monitoring}
\bibfield{author}{\bibinfo{person}{Oded Maler} {and} \bibinfo{person}{Dejan Nickovic}.} \bibinfo{year}{2004}\natexlab{}.
\newblock \showarticletitle{Monitoring temporal properties of continuous signals}.
\newblock In \bibinfo{booktitle}{\emph{Formal Techniques, Modelling and Analysis of Timed and Fault-Tolerant Systems}}. \bibinfo{publisher}{Springer}, \bibinfo{pages}{152--166}.
\newblock


\bibitem[Mazurek et~al\mbox{.}(2023)]%
        {mazurek:micro:2023}
\bibfield{author}{\bibinfo{person}{Filip Mazurek}, \bibinfo{person}{Arya Tschand}, \bibinfo{person}{Yu Wang}, \bibinfo{person}{Miroslav Pajic}, {and} \bibinfo{person}{Daniel~J. Sorin}.} \bibinfo{year}{2023}\natexlab{}.
\newblock \showarticletitle{Rigorous Evaluation of Computer Processors with Statistical Model Checking}. In \bibinfo{booktitle}{\emph{Proceedings of the 56th IEEE/ACM International Symposium on Microarchitecture}}.
\newblock


\bibitem[Purnal et~al\mbox{.}(2021)]%
        {purnal:sp:2021}
\bibfield{author}{\bibinfo{person}{Antoon Purnal}, \bibinfo{person}{Lukas Giner}, \bibinfo{person}{Daniel Gruss}, {and} \bibinfo{person}{Ingrid Verbauwhede}.} \bibinfo{year}{2021}\natexlab{}.
\newblock \showarticletitle{Systematic analysis of randomization-based protected cache architectures}. In \bibinfo{booktitle}{\emph{2021 IEEE Symposium on Security and Privacy (SP)}}. IEEE, \bibinfo{pages}{987--1002}.
\newblock


\bibitem[Saileshwar and Qureshi(2021)]%
        {saileshwar:usenix-security:2021}
\bibfield{author}{\bibinfo{person}{Gururaj Saileshwar} {and} \bibinfo{person}{Moinuddin Qureshi}.} \bibinfo{year}{2021}\natexlab{}.
\newblock \showarticletitle{{MIRAGE}: Mitigating {Conflict-Based} Cache Attacks with a Practical {Fully-Associative} Design}. In \bibinfo{booktitle}{\emph{30th USENIX Security Symposium (USENIX Security 21)}}. \bibinfo{pages}{1379--1396}.
\newblock


\bibitem[Saileshwar and Qureshi(2019)]%
        {saileshwar:micro:2019}
\bibfield{author}{\bibinfo{person}{Gururaj Saileshwar} {and} \bibinfo{person}{Moinuddin~K. Qureshi}.} \bibinfo{year}{2019}\natexlab{}.
\newblock \showarticletitle{CleanupSpec: An "Undo" Approach to Safe Speculation}. In \bibinfo{booktitle}{\emph{Proceedings of the 52nd Annual IEEE/ACM International Symposium on Microarchitecture}}. \bibinfo{pages}{73–86}.
\newblock


\bibitem[Seznec(1993)]%
        {seznec:isca:1993}
\bibfield{author}{\bibinfo{person}{Andr\'{e} Seznec}.} \bibinfo{year}{1993}\natexlab{}.
\newblock \showarticletitle{A case for two-way skewed-associative caches}. In \bibinfo{booktitle}{\emph{Proceedings of the 20th International Symposium on Computer Architecture}}.
\newblock


\bibitem[Tromer et~al\mbox{.}(2010)]%
        {tromer:jcrypto:2010}
\bibfield{author}{\bibinfo{person}{E. Tromer}, \bibinfo{person}{D.A. Osvik}, {and} \bibinfo{person}{A. Shamir}.} \bibinfo{year}{2010}\natexlab{}.
\newblock \showarticletitle{Efficient Cache Attacks on AES, and Countermeasures}.
\newblock \bibinfo{journal}{\emph{Journal of Cryptology}}  \bibinfo{volume}{23} (\bibinfo{year}{2010}), \bibinfo{pages}{37--71}.
\newblock


\bibitem[Wang et~al\mbox{.}(2019)]%
        {wang_tecs19}
\bibfield{author}{\bibinfo{person}{Yu Wang}, \bibinfo{person}{Mojtaba Zarei}, \bibinfo{person}{Borzoo Bonakdarpour}, {and} \bibinfo{person}{Miroslav Pajic}.} \bibinfo{year}{2019}\natexlab{}.
\newblock \showarticletitle{Statistical Verification of Hyperproperties for Cyber-Physical Systems}.
\newblock \bibinfo{journal}{\emph{ACM Trans. Embed. Comput. Syst.}} \bibinfo{volume}{18}, \bibinfo{number}{5s}, Article \bibinfo{articleno}{92} (\bibinfo{date}{Oct.} \bibinfo{year}{2019}), \bibinfo{numpages}{23}~pages.
\newblock
\showISSN{1539-9087}
\href{https://doi.org/10.1145/3358232}{doi:\nolinkurl{10.1145/3358232}}


\bibitem[Werner et~al\mbox{.}(2019)]%
        {werner:css:2019}
\bibfield{author}{\bibinfo{person}{Mario Werner}, \bibinfo{person}{Thomas Unterluggauer}, \bibinfo{person}{Lukas Giner}, \bibinfo{person}{Michael Schwarz}, \bibinfo{person}{Daniel Gruss}, {and} \bibinfo{person}{Stefan Mangard}.} \bibinfo{year}{2019}\natexlab{}.
\newblock \showarticletitle{SCATTERCACHE: thwarting cache attacks via cache set randomization}. In \bibinfo{booktitle}{\emph{Proceedings of the 28th USENIX Conference on Security Symposium}}. \bibinfo{pages}{675–692}.
\newblock


\bibitem[Yang et~al\mbox{.}(2023)]%
        {yang:isca:2023}
\bibfield{author}{\bibinfo{person}{Yuheng Yang}, \bibinfo{person}{Thomas Bourgeat}, \bibinfo{person}{Stella Lau}, {and} \bibinfo{person}{Mengjia Yan}.} \bibinfo{year}{2023}\natexlab{}.
\newblock \showarticletitle{Pensieve: Microarchitectural Modeling for Security Evaluation}. In \bibinfo{booktitle}{\emph{Proceedings of the 50th Annual International Symposium on Computer Architecture}}.
\newblock


\bibitem[Zarei et~al\mbox{.}(2020a)]%
        {zarei_hscc20}
\bibfield{author}{\bibinfo{person}{Mojtaba Zarei}, \bibinfo{person}{Yu Wang}, {and} \bibinfo{person}{Miroslav Pajic}.} \bibinfo{year}{2020}\natexlab{a}.
\newblock \showarticletitle{Statistical Verification of Learning-Based Cyber-Physical Systems}. In \bibinfo{booktitle}{\emph{Proceedings of the 23rd International Conference on Hybrid Systems: Computation and Control}} (Sydney, New South Wales, Australia) \emph{(\bibinfo{series}{HSCC'20})}. Article \bibinfo{articleno}{12}, \bibinfo{numpages}{7}~pages.
\newblock
\showISBNx{9781450370189}
\href{https://doi.org/10.1145/3365365.3382209}{doi:\nolinkurl{10.1145/3365365.3382209}}


\bibitem[Zarei et~al\mbox{.}(2020b)]%
        {zarei:hscc:2020}
\bibfield{author}{\bibinfo{person}{M. Zarei}, \bibinfo{person}{Y. Wang}, {and} \bibinfo{person}{M. Pajic}.} \bibinfo{year}{2020}\natexlab{b}.
\newblock \showarticletitle{Statistical Verification of Learning-Based Cyber-Physical Systems}. In \bibinfo{booktitle}{\emph{{ACM} International Conference on Hybrid Systems: Computation and Control}}.
\newblock


\bibitem[Zhang et~al\mbox{.}(2013)]%
        {zhang:hasp:2013}
\bibfield{author}{\bibinfo{person}{Tianwei Zhang}, \bibinfo{person}{Si Chen}, \bibinfo{person}{Fangfei Liu}, {and} \bibinfo{person}{Ruby~B. Lee}.} \bibinfo{year}{2013}\natexlab{}.
\newblock \showarticletitle{Side Channel Vulnerability Metrics: The Promise and the Pitfalls}. In \bibinfo{booktitle}{\emph{Proceedings of the 2nd International Workshop on Hardware and Architectural Support for Security and Privacy}}.
\newblock


\end{thebibliography}




\end{document}